\documentclass[12pt,english, A4paper, onecolumn, draftcls]{IEEEtran}
\usepackage[latin1]{inputenc}

\usepackage{babel}


\usepackage[T1]{fontenc}
\usepackage{array}
\usepackage{graphicx}
\usepackage[caption=false,font=footnotesize]{subfig}
\usepackage{dsfont}
\usepackage{fixltx2e}
\usepackage{cite}
\usepackage{algorithm}
\usepackage{algorithmic}
\usepackage{xfrac}
\usepackage{wrapfig}
\usepackage[cmex10]{amsmath}
\usepackage{amsfonts,amssymb}
\usepackage{booktabs}
\usepackage{color}
\usepackage{epstopdf}
\usepackage{float}
\usepackage[T1]{fontenc}
\usepackage{multicol, blindtext}
\usepackage{mathtools}
\usepackage{url}	
\usepackage{hyperref}
\usepackage{tikz}
\definecolor{gold}{rgb}{0.85,.66,0}

\newcommand{\colk}{\textcolor{black}}

\newcommand{\diag}{{\rm diag}}

\usepackage{soul}  

\begin{document}
\title{Efficient ZF-WF Strategy for Sum-Rate Maximization of MU-MISO Cognitive Radio Networks}
\author{Lucas~Claudino and Taufik~Abrao	
\thanks{L. Claudino and T. Abrao are with Department of Electrical Engineering of Londrina State University, Parana BR. Email: lsclaudino@gmail.com\,\, taufik@uel.br}}

\maketitle

\begin{abstract}
This article presents {an efficient quasi-optimal sum rate (SR)} maximization technique based on zero-forcing water-filling (ZFWF) algorithm directly applied to cognitive radio networks (CRNs). {We have defined the non-convexity nature of the optimization problem in the context of CRNs while we have offered all necessary conditions to solve the related {\colk{SR}} maximization problem, which considers power limit at cognitive transmitter and interference levels at primary users (PUs) and secondary users (SUs)}.  {A general expression capable to determine the optimal number of users as a function of the main system parameters, namely the signal-to-interference-plus-noise ratio (SINR) and the number of BS antennas is proposed. Our} numerical results for the CRN performance are analyzed in terms of {both BER and sum-capacity for the proposed ZF-WF precoding technique, and compared to the classical minimum mean square error (MMSE), corroborating the effectiveness of the proposed technique operating in multi user multiple input single output (MU-MISO) CRNs.}
\end{abstract}


\IEEEpeerreviewmaketitle

\section{Introduction}
\label{Sec_Introduction}
The spectrum is a limited resource that, until nowadays, has been regulated in a \textit{fixed spectrum access} form. This means that each sub-band of the total spectrum bandwidth is assigned to one specific owner (PU) who has paid for the right to transmit over {these} frequencies; indeed, no other user is allowed to exploit this preallocated spectrum, regardless if the PU is using it or not. The problem is that, in the past decades, wireless technologies have been significantly developed and these fixed frequency bandwidths are becoming scarce. In recent studies, regulatory commissions, {such as} Federal Communications Commission (FCC), have discovered that the spectrum is underutilized \cite{FCC_02_135,Datla_2009}. These studies reveal the need for new {and more efficient} schemes of spectrum {management}. Hence, researchers have been looking for strategies to enhance spectrum utilization efficiency. A recent technology, known as cognitive radio (CR) aims to solve the {recurrent and critical} spectrum scarcity problem via proposing a wireless transceiver able to interact with the environment and change its transmission parameters in order to achieve a better performance \cite{FCC_03_222}.

The concept of CR has been firstly introduced in \cite{Mitola_1999}, where the author stated that CR may be interpreted as an evolution of software defined radio (SDR), where the various SDRs present a high level of computational intelligence. Such intelligence makes them able to mimic some human cognitive behavior like observation, orientation, planning, decision and action, in order to derivate a broad view about the wireless scenario and provide appropriate wireless services.

The CR is basically a system with high environmental awareness able to dynamically access all available bandwidth. Therefore, a CR is a special radio system with two main abilities: the cognition capability and the reconfigurability \cite{Khattab_2013_CRNbook}.  Cognition of a CR is basically the ability to sense the environment and observe the spectral opportunities so the radio is able to identify the available spectrum bands. Reconfigurability is related to the fact t{hat a CR, after estimating the bandwidth usage, is able to interactively adapt its transmissions values and plans in terms of power, bandwidth and time {availability}.

A typical CRN layout consists of a series of PUs coexisting harmoniously with the CR devices, namely secondary users. PUs are also known as licensed users, which are the ones who own the license to transmit over some specific bandwidth. CR basically proposes that SUs operate over the bandwidth, even though they do not hold a license. In order to do so, a series of constraints must be followed, {\it i.e.}, the SU may only operate when the PU is not transmitting or, in case of PU activity, the SU must not overcome an energy threshold in order not to affect the PU's transmission \cite{Khattab_2013_CRNbook}. 

\colk{Some work has been done in CR scenarios regarding the SR maximization. This optimization problem is treated as a tradeoff between spatial multiplexing at SUs and interference avoidance at PUs in \cite{Zhang2008}. Also, authors in \cite{Zhang2008} propose sub-optimal SVD-based algorithms (Singular Value Decomposition) to maximize the sum capacity of secondary transmissions. The work in \cite{LZhang2009} proposes a weighted \colk{SR} problem with solution based on iterative subgradient algorithms, once the resultant relaxed problem can not be solved with traditional iterative water-filing techniques. Recently, a relaxed problem is proposed in \cite{Zheng2014}, where all constraints are transformed into a nonnegative matrix spectral radius. This relaxation is then solved with polynomial-time iterative algorithms. Also, physical quantities are analyzed and taken into account, \textit{i.e.} channel parameters, transmission power and achievable rates for SUs.}

\colk{Recently, a linear precoder design for an underlay cognitive radio MIMO broadcast channel with multiuser interference elimination provided by zero-forcing is proposed in \cite{VanDinh2016}. To develop an efficient precoder design for multiuser MIMO-CRN under interference constraints, the authors firstly apply a rank relaxation method to transform the problem into a convex problem, and then deploy a barrier interior-point method to solve the resulting saddle point problem. Solving a system of discrete-time Sylvester equations, authors demonstrated with numerical results a substantial complexity reduction compared to conventional methods.} \colk{Considering the multiuser interference alignment (IA) technique, the work \cite{He.2016} analyses the problem of SINR decreasing due to channel conditions in IA-based CRNs, which reduce the quality of service (QoS) of PUs. In this context, the authors propose a multiuser-diversity-based IA scheme applicable to CRNs. Under small number of SUs, the authors have found that the IA network can accommodate all the users simultaneously without mutual interference; however, under large number of SUs, the IA-based CRN is not effective, being not able to accommodate all the PUs and SUs simultaneously with perfect elimination of interferences.}

\colk{Recent work has dealt with MIMO-CR downlink architectures and developed a block matrix strategy to cancel interference between SUs and to PUs to keep all interference levels under a certain threshold \cite{Turki2015}. The authors use the second order Karush-Kuhn-Tucker (KKT) conditions to design a precoder able to deal with this interference. Also, the sum rate maximization problem in CRN is examined when imperfect channel state information (CSI) is available, and come up with high computationally complex but optimal power distribution scheme.}

\colk{The \colk{SR} maximization problem of CRNs has been a classical problem in wireless system design \cite{Zhang.2008, Kim.2011}. Specifically,  in \cite{Zhang.2008} it was considered the weighted \colk{SR} maximization problem for CRNs with multiple-SUs MIMO broadcast channel under sum power constraint and interference power constraints. Authors have shown that optimization problem is a nonconvex problem, but can be transformed into an equivalent convex MIMO multiple access channel problem. Moreover, the work in \cite{Kim.2011} deals with the optimal resource allocation problem in MIMO Adhoc CRNs; a semi distributed algorithm and a centralized algorithm based on geometric programming and network duality were introduced under the interference constraint at primary receivers, aiming obtaining a locally optimal linear precoder for the nonconvex weight \colk{SR} maximization problem. It is worth to note that these SR maximization algorithms in CRNs generally result in excessive computational complexity combined to slow convergence.}
 
\colk{This contribution} is devoted to analyze an underlay MU-MISO CRN, where all SUs are equipped with a single antenna and communicate with a multiple-antenna base-station (BS). The goal is to design beamforming vectors aiming to maximize the SU's sum {rate} while reducing (or even avoiding) interference levels seen at all PUs. The {\it main contribution} of this paper consists in {\it combining low-complexity power allocation optimization design with a conventional precoding solution aiming to alleviate the constraints requirements} directly applicable to the maximization of CRNs \colk{SR} capacity. {We have provided a general fitting expression capable to determine the optimal number of users as a function of the main system parameters SINR and number of BS antennas} 

The work is divided as follows. Section \ref{Sec_System} models the CRN system scenarios, explaining basics of precoding techniques deployed in this contribution. Section \ref{Sec_Convexity} states the optimization problem and analyzes its convexity based on non-linear optimization theory and KKT necessary conditions. Section \ref{Sec_ZFWF} uses the well-known zero-forcing (ZF) precoding technique to reduce the constraints and narrow down the problem to a power allocation optimization problem. For this combined strategy, we have provided a comprehensive analysis and details on the design when applied to the CRNs. Moreover, corroborative numerical results and respective analysis are presented in section \ref{Sec_Results}, demonstrating the improvement offered by the proposed combined optimization strategy in terms of sum capacity of an entire secondary network constrained by interference limits to PUs. Additionally, our numerical results also emphasize system capacity improvements upon other classical beamformer strategies, which are not designed to maximize capacity or do not intend to cancel (or alleviate) interference to PUs. {Final} remarks and future work are {offered} in section \ref{Sec_Conclusion}. 

To facilitate the readability of the paper, in the following we provide the notation and a list of symbols adopted in this work.

\noindent \textbf{Notation}:  {$x$} represents a scalar variable, while $ \mathbf{x} $ is a vector and $ \mathbf{X} $ is a matrix. Hermitian matrix {is denoted by} $ (\cdot)^H $. $ \nabla f $ is the gradient of $ f $ and $ \nabla^2f $ is {the respective} Hessian matrix.\\

\begin{tabular}{ll}
$ K $:				&	Number of SUs\\
{$ K^* $:}		&	{Optimum number of SUs}\\
$ \mathcal{K} $: 	& 	Set of SUs\\
$ M $:				&	Number of PUs\\
$ \mathcal{M} $:	& 	Set of PUs\\
$ \mathcal{S} $: 	& 	Set of active SUs\\
$ n_\textsc{bs} $:	& 	Base station's number of antennas\\
$ I_m $:			&	Interference limit to $ m $-th PU\\	
$ I_p $:			&	Interference from PUs to SUs\\
$ P_\textsc{bs} $	& 	SU-BS's power constraint.\\
$ \mathcal{C}_k $	& 	Capacity of user $ k $.\\
$ \varphi, \,\beta $:	&	{Angular, linear coefficients of the linear fitting}\\
$ y_k:$ 			& 	Received signal.\\
$ \eta_k $:			& 	AWGN noise.\\
\colk{$ n_k $}:			& 	\colk{AWGN noise plus constant interference from PUs}\\
$ {\rm x}_k$,\, $\mathbf{x}$:	& 	Transmitted symbol (scalar) and signal vector\\
\colk{$ \mathbf{z}_j $}	&	\colk{PU's transmitted signal}\\
$ \mathbf{h}_k$, $ \mathbf{g}_m $, \colk{$ \mathbf{q}_{m,k} $}:&	BS-SU, BS-PU, \colk{PU-SU} link's channel vectors\\
$ {\bf w}_k$, ${\bf t}_k $:& 	General and ZF precoding vectors\\
$ \mathbf{p} $:		& 	Power allocation vector\\
$ \mathbf{H}$, $\mathbf{G}$: & SU and PU channel matrices\\
$ \mathbf{W} $:		& 	Precoding matrix\\
$ \mathbf{F}_{-k} $:& 	PU and SU channel matrices concatenated, except user $k$.\\
\end{tabular}

\section{System model}
\label{Sec_System}
In this article a MU-MISO underlay\footnote{Remember from the definition of underlay CR that any SINR  measurement at PUs must be below a pre-determined threshold \cite{Biglieri_BookPrincCR}.} A CR system is considered, where $ K $ single antenna SUs are simultaneously transmitting with $ M $ single-antenna PUs over the same frequency bandwidth. As illustrated in Fig. \ref{fig_scenario}, each $ k $-th link between BS and $ k $-th SU has a channel response $ \mathbf{h}_k\in \mathbb{C}^{n_\textsc{bs}\times 1}$, $ k\in\mathcal{K}=\{1,\,,2,  \dots, \,K\} $, $\mathbf{g}_m \in \mathbb{C}^{n_\textsc{bs}\times 1} $, $ m\in \mathcal{M}=\{1,\, ,2,\dots, \, M\} $ is the channel matrix for the $ m $-th BS-PU link, which is considered a form of interference \colk{for any PU. Also, and all PUs are considered to be constantly transmitting, their transmission signal is seen as interference at SUs; consequently, there is a channel vector $ \mathbf{q}_{m,k} $ relating each $ m $-th PU and $ k $-th SU.}

\begin{figure}[!htbp]
	\centering
	\includegraphics[width=0.4\textwidth]{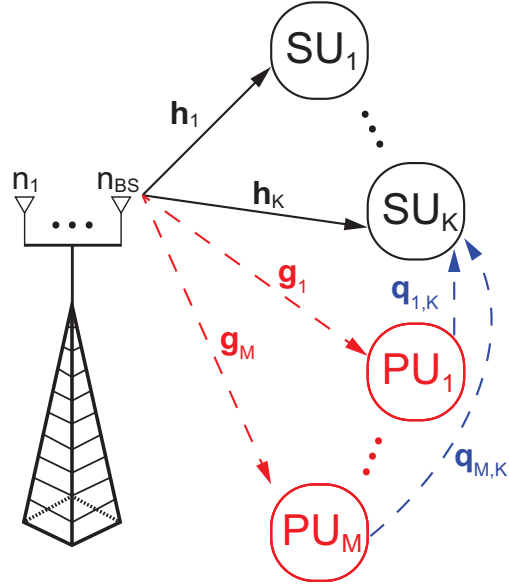}
	\vspace{-2mm}
	\caption{Typical MISO CRN scenario.}
	\label{fig_scenario}
\end{figure}

In such scenario, the secondary BS is responsible for choosing each $ k $-th link's appropriate transmit power in order to keep all interference power under an upper limit $ I_m $, which varies for each $ m $-th PU. The goal is to maximize the sum capacity via designing an optimal SU precoding vector $ {\bf w}_k $ attending to per-antenna and interference power constraints. Also, capacity depends on the interference level from PUs to SUs \colk{(related to channel vectors $ \mathbf{q_{m,k}} $), which will later be considered, for computational simplicity, a constant with average power denoted by $ I_p $} to be added to the denominator of the SINR expression.

Ideally, precoding relies on perfect CSI; hence, knowledge of $ \mathbf{h}_k $ at BS is assumed. Even though this assumption is an ideal hypothesis, the problem of imperfect CSI can be neglected with the purpose of investigating the potential of new transmission schemes combining resource allocation and precoding techniques. The perfect CSI assumption has also been considered in recent contributions, including \cite{VanDinh2016,HieuNguyen2016,YYHe2014,LGallo2011,LZhang2009} in order to ease studies of beamforming design for CR-MU-MIMO networks.

A downlink (DL) transmission is considered, where a baseband signal $ \mathbf{x} $ contains all transmitted symbols $ {\rm x}_k $ and beamforming vector $ \mathbf{w}_k \in \mathbb{C}^{n_\textsc{bs}\times 1}$ associated to every $ k\in \mathcal{K} $ SU user:
\begin{equation}
\mathbf{x} = \sum_{k\in \mathcal{K}}\mathbf{w}_k{\rm x}_k
\label{eq_x}
\end{equation}

Received signal $ {\rm y}_k $ at $ k$-th SU is then expressed as a function of the signal destined to user $ k $ plus an interference from  another secondary transmissions \colk{and interference from PUs}:
\begin{equation}
{\rm y}_k = \mathbf{h}_k^H\mathbf{w}_k{\rm x}_k + \sum_{i\in\mathcal{K},i\neq k}\mathbf{h}_k^H\mathbf{w}_i{\rm x}_i + \colk{\sum_{j=1}^M\mathbf{q}_{j,k}^H\mathbf{z}_j} + \eta_k
\label{eq_y}
\end{equation}
where $ \eta_k \sim \mathcal{CN}(0,\sigma_k^2)$ is the additive white Gaussian noise (AWGN) and transmitted symbols have normalized power $ {\rm x}_k \sim \mathcal{CN}\left(0,1\right) $; so, transmitted power is set only by precoding vectors. \colk{Also, there is a transmitted signal $ \mathbf{z}_j $ for all $ M $ PUs; however, SUs do not have any information about primary user characteristics or its transmitted symbols.}

\colk{As mentioned before, all PUs are considered to be constantly transmitting. This is an unavoidable matter, once SUs do not have any power upon primary transmissions and must design strategies to lead with this interference. Considering that all $ M $ PUs transmit over the entire period of time, for sake of simplicity, an average interference power $ I_p $ can be used instead of calculating $ \mathbb{E}\left[\sum_{j=1}^M\left(\mathbf{q}_{j,k}^H\mathbf{z}_j\right)^2\right] $. Hence, once this constant power interference has been considered, $I_p$ can be incorporated into the noise term in \eqref{eq_y} as $n_k$, such that $n_k \sim \mathcal{CN}\left(0,\sigma_k^2+I_p\right)$. As a result, the received signal at $ k $-th SU can be re-written as:
\begin{equation}\label{eq_y_2}
{\rm y}_k = \mathbf{h}_k^H\mathbf{w}_k{\rm x}_k + \sum_{i\in\mathcal{K},i\neq k}\mathbf{h}_k^H\mathbf{w}_i{\rm x}_i + {n_k}
\end{equation}}
Given the received signal in \eqref{eq_y_2} and perfect channel knowledge at transmitter's side, it is possible to design a precoder to optimize the received signal in terms of power, signal integrity, QoS, bit error rate (BER) or capacity. The optimization metric chosen in this work is the sum capacity of the CRN. Indeed, the first part in this work focuses on the analysis and comparison of different precoders in terms of sum capacity.

\section{Convexity analysis}
\label{Sec_Convexity}
The optimization problem consists of a \colk{SR} maximization with respect to all SUs in the CRN, constrained by power limit at BS, $ P_\textsc{bs},$ and a maximum interference to any PU transmission, $ I_m$. For this case scenario, every $ k $-th SU transmission is subject to interference from others SUs and additive noise. Hence, the SINR is defined as:
\begin{equation}\label{eq:SINR_k}
\gamma_k = \dfrac{\mathbf{h}_k \mathbf{w}_k \mathbf{w}_k^H\mathbf{h}_k^H}{\sum_{j\neq k}^{K}\mathbf{h}_k\mathbf{w}_j\mathbf{w}_j^H\mathbf{h}_k^H+\sigma^2_k  + I_p} \quad \forall k\in\mathcal{K}
\end{equation}

The \colk{SR} maximization problem subject to power and interference constraints is then defined as:
\begin{subequations}
	\begin{align}
	\max_{\{\mathbf{w}_1,\dots,\mathbf{w}_K\}}& \quad \sum_{k=1}^{K}\log_2 \left(1+\gamma_k\right) \label{Eq_Optz_1_Cost}\\
		{\rm s.t:}& \quad \mathbf{w}_k^H\mathbf{w}_k \leq P_\textsc{bs}, \,\, \forall k\in\mathcal{K} \label{Eq_Optz_1_Ct1}\\
		&\quad \sum_{k\in\mathcal{K}}\mathbf{g}_m\mathbf{w}_k\mathbf{w}_k^H\mathbf{g}_m^H\leq I_m, \,\, \forall m\in\mathcal{M} \label{Eq_Optz_1_Ct2}
	\end{align}
\label{eq_optz_1}
\end{subequations}

The problem in \eqref{eq_optz_1} is said to be convex if both cost function and inequalities constraints are convex. A few works say this is a non-convex optimization problem, however they do not prove it \cite{LGallo2011,YYHe2014}. Constraints \eqref{Eq_Optz_1_Ct1} and \eqref{Eq_Optz_1_Ct2} are both quadratic functions with domain $ \mathbb{R}^\textsc{K}\rightarrow \mathbb{R}$, which are known to be convex \cite{Boyd_book}. Hence, the problem is reduced to identifying whether \eqref{Eq_Optz_1_Cost} is convex or not. Via composition property, given an arbitrary function $ f(x)=h(g(x)) $ is concave if $ g(x) $ and $ h(x) $ are concave and non-decreasing. Logarithmic functions are concave non-decreasing, which brings us to analyze the concavity of the $ \gamma_k $ expression.

A function is convex/concave if its Hessian matrix is positive/negative semidefinite. The multidimensional analysis of \eqref{Eq_Optz_1_Cost} is quite complex; however, if the unidimensional case is proved to be non-concave, the cost function is non-concave for any dimension. In contrast, if the unidimensional case is concave, no further assumptions can be made about the multidimensional one. Let us assume:
\begin{equation}
f(w_k) = \dfrac{h_k w_k w_k^* h_k^*}{\sum_{j\neq k}^{K} h_k w_j w_j^* h_k^*+\sigma^2_k + I_p}
\end{equation}
then, $ f(w_k) $ is concave if and only if the Hessian matrix $\mathbf{H}_f$ is negative semidefinite:
\begin{equation}
\mathbf{H}_f = \nabla^2f \triangleq \begin{bmatrix}
\dfrac{\partial^2 f}{\partial w_1^2}  & \cdots & \dfrac{\partial^2 f}{\partial w_1 w_K}	\\
\vdots 								  & \ddots & \vdots									\\
\dfrac{\partial^2 f}{\partial w_K w_1}& \cdots & \dfrac{\partial^2 f}{\partial w_K^2}
\end{bmatrix}\preceq \mathbf{0}
\end{equation}
The partial derivatives with respect to the main diagonal of $ \mathbf{H}_f $ are:
\begin{align}
\dfrac{\partial^2 f}{\partial w_k^2} &= \dfrac{\partial}{\partial w_k}\left[ \dfrac{2\left|h_k\right|^2w_k}{\sum_{j\neq k}\left|h_k\right|^2\left|w_j\right|^2+\sigma_k^2 + I_p} \right] \nonumber\\
	&= \dfrac{2\left|h_k\right|^2}{\sum_{j\neq k}\left|h_k\right|^2\left|w_j\right|^2+\sigma_k^2 + I_p}
\end{align}
which are all non-negative values.

Additionally, the partial derivatives with respect to the off-diagonal elements are:
\begin{align}
\dfrac{\partial^2 f}{\partial w_kw_j} &= \dfrac{\partial}{\partial w_j}\left[ \dfrac{2\left|h_k\right|^2w_k}{\sum_{j\neq k}\left|h_k\right|^2\left|w_j\right|^2+\sigma_k^2 + I_p} \right] \nonumber\\
&= 2\left|h_k\right|^2w_k\dfrac{-2\sum_{j\neq k}h_kw_jh_k}{\left(\sum_{j\neq k}\left|h_k\right|^2\left|w_j\right|^2+\sigma_k^2 + I_p\right)^2}\nonumber\\
&= -\dfrac{4\left|h_k\right|^4w_k\sum_{j\neq k}w_j}{\left(\sum_{j\neq k}\left|h_k\right|^2\left|w_j\right|^2+\sigma_k^2 + I_p\right)^2}
\label{eq_optz_1_NonDiag}
\end{align}

Analysis of \eqref{eq_optz_1_NonDiag} shows that the Hessian matrix is non symmetric:
\begin{align}
\dfrac{\partial^2 f}{\partial w_1w_2} &= -\dfrac{4\left|h_1\right|^4w_1\sum_{j\neq 1}w_j}{\left(\sum_{j\neq 1}\left|h_1\right|^2\left|w_j\right|^2+\sigma_1^2 +  I_p\right)^2} \nonumber\\
&\neq \dfrac{\partial^2 f}{\partial w_2w_1} = -\dfrac{4\left|h_2\right|^4w_2\sum_{j\neq 2}w_j}{\left(\sum_{j\neq 2}\left|h_2\right|^2\left|w_j\right|^2+\sigma_2^2 + I_p\right)^2}
\end{align}

A simple property is that, a negative semidefinite matrix has all its eigenvalues smaller or equal to zero. If a matrix is not symmetric, then its eigenvalues are not necessarily in $ \mathbb{R} $; hence, this matrix is not negative semidefinite \cite{SemidefMatrixBook}.

Additionally, numerical simulations aiming to corroborate this fact have been proceeded. A set of matrices with negative main diagonal and complex-normally distributed off-diagonal elements has been generated. Indeed, some of such matrices showed to have both positive and negative eigenvalues, which corroborates that the Hessian matrix of $\gamma_k$, Eq. \eqref{eq:SINR_k}, is not negative semidefinite.

\subsection{KKT necessary conditions and absence of closed expression}
\label{Sec_KKT}
Convex problems can be straightforwardly solved applying KKT necessary conditions. Indeed, a possible optimal point $ \mathbf{w}_k^* $ would be found setting the derivative of the Lagrangian of the problem \eqref{eq_optz_1}  to zero, where the Lagrangian is defined in Eq. \eqref{eq_Lagrangian}.

\begin{equation}
\mathcal{L}(\mathbf{w}_k,\lambda_k,\mu_m) = \sum_{k=1}^{K}\log_2(1+{\rm SINR_k})-\sum_{k=1}^{K}\lambda_k\left( \dfrac{\mathbf{w}_k\mathbf{w}_k^H}{P_\textsc{BS}} -1 \right) - \sum_{m=1}^{M}\dfrac{\mu_m}{I_m} \left( \sum_{k=1}^{K} \mathbf{g}_m\mathbf{w}_k\mathbf{g}_m^H\mathbf{w}_k^H -1 \right)
\label{eq_Lagrangian}
\end{equation}

Similar to \cite{LGallo2011}, gradient of the Lagrangian can be set to zero as in Eq. \eqref{eq_Lagrangian_diff}, where $ \lambda_k $ and $ \mu_m $ are the Lagrange multipliers. Inspection of \eqref{eq_Lagrangian_diff} shows that all three therms depend on the optimization variable $\bf w$, which makes it impractical to find a close expression for the optimum precoding vector solution. Hence, herein we prefer to elaborate iterative methods to find near-optimum solutions.
\begin{align}
\dfrac{\partial \mathcal{L}(\mathbf{w}_k,\lambda_k,\mu_m) }{\partial \mathbf{w}_k^*} &= \dfrac{q_k^*\mathbf{h}_k^H}{\ln(2)} - \sum_{j\neq k}\dfrac{q_j\omega_jq_j^*\mathbf{h}_j^H\mathbf{h}_j\mathbf{w}_k}{\ln(2)} - \dfrac{\lambda_k\mathbf{w}_k}{P_\textsc{bs} - \sum_{m=1}^{M}\dfrac{\mu_m}{I_m} \mathbf{g}_m^H\mathbf{g}_m\mathbf{w}_k} = \mathbf{0} \label{eq_Lagrangian_diff}\\
\text{where} \quad q_k &= e_kd_k^{-1}\mathbf{w}_k^H\mathbf{h}_k^H, \quad  e_k = \left(1 + \mathbf{h}_k\mathbf{w}_k\mathbf{w}_k^H\mathbf{h}_k^Hd_k^{-1}\right) = {\omega_k^{-1}}, \quad
d_k = \sum_{j\neg k,j\in \mathcal{K}}\mathbf{h}_k\mathbf{w}_j\mathbf{w}_j^H\mathbf{h}_k^H \nonumber
\end{align}

\section{Zero-forcing water filling precoding}
\label{Sec_ZFWF}
It this section a mixed technique known as ZFWF precoding is analyzed for MU-MISO systems. ZF is largely applied to MU-MISO networks due to its facility of design beamforming vectors for the $k$th user, $\mathbf{t}_k$, such that users receive interference free signals due to orthogonality between beamforming vectors of different users. In CR scenarios, ZF is able to provide a design that eliminates interference between distinct SUs. Herein, a suboptimal ZF solution for problem \eqref{eq_optz_1} is considered as a strategy of interference canceling for both classes of users, SUs and PUs. 

The beamforming vector is divided into power allocation ($p_k$) and interference cancellation ($\mathbf{t}_k$) parts: 
\begin{equation}
\mathbf{w}_k = \sqrt{p_k}\mathbf{t}_k 
\end{equation}

Interference canceler vector $ \mathbf{t}_k $ is designed such that it is simultaneously orthogonal to the $i$th SU and $m$th PU channel vectors: 
 \begin{subequations}
\begin{align}
 \mathbf{h}_i^H \mathbf{t}_k=0, & \, \qquad \forall i,k \in \mathcal{K}, \quad \, i\neq k \\
 \mathbf{g}_m^H \mathbf{t}_k=0, & \, \qquad \forall m \in \mathcal{M} 
\end{align}
\end{subequations}

Let us concatenate all SU and PU channel vectors, except the $k$th SU channel vector, as a matrix:
$ \mathbf{F}_{-k} \triangleq \left\{ \mathbf{g}_1,\dots,\mathbf{g}_M, \mathbf{h}_1,\dots, \mathbf{h}_{k-1},\mathbf{h}_{k+1},\dots,\mathbf{h}_K \right\} \in \mathbb{C}^{n_\textsc{bs}\times (M+K-1)}$. The interference free constraint is then re-written as $ \mathbf{F}_{-k}^H\mathbf{t}_k=\mathbf{0} $; indeed, $ \mathbf{t}_k $ should be designed to lie on the null-space of $ \mathbf{F}_{-k} $. This assumption will simplify the original problem in \eqref{eq_optz_1} such that \eqref{Eq_Optz_1_Ct2} is eliminated, once the beamforming vector design guarantees zero interference from SUs to PUs. Additionally, $ \mathbf{p} = \left[p_1,\dots,p_k\right] $ is solution to the simplified decoupled power allocation problem. Even though the original problem was proved to be non-convex, the interference-free constraint imposed by the ZFWF precoder simplifies the SINR expression and, from \eqref{eq_optz_1_NonDiag}, setting the multiplication $ w_jh_k=0, \, \forall \, j\neq k$, all off-diagonal elements become zero. Indeed, the Hessian matrix has its main diagonal elements greater or equal zero and the off-diagonal equal zero, what is characteristic of a positive semidefinite matrix.

Under such assumptions, the optimization problem is simply re-written as:
\begin{subequations}
	\begin{align}
	\max_{\{\mathbf{t}_k\},\mathbf{p}} \quad &\sum_{k=1}^{K} \log_2\left(1+\gamma_k\right) \\
	{\rm s.t}:	\quad&\sum_{k=1}^{K}p_k\mathbf{t_k}^H\mathbf{t}_k \leq P_\textsc{bs}, \, \forall k\in\mathcal{K}\\
			&\mathbf{F}_{-k}^H\mathbf{t}_k = \mathbf{0}
	\end{align}
\end{subequations}
\begin{enumerate}
	\item When $(M+K-1)< n_\textsc{bs}$, $ {\rm rank} \left(\mathbf{F}_{-k}\right)<n_\textsc{bs}$; consequently, $ \mathbf{F}_{-k}^H\mathbf{t}_k =\mathbf{0}, \forall k$ presents an infinite number of solutions, including the optimal $ \mathbf{W}^* =\mathbf{T}\cdot\diag\left(\sqrt{\mathbf{p^*}}\right)$, where $ \mathbf{T} $ is the classical ZF solution: $ \mathbf{T}^* = \mathbf{T}'\left(\mathbf{T}'\mathbf{T}'^H\right)^{-1}$, where $ \mathbf{T}' = \left(\mathbf{I} - \mathbf{G}^H\mathbf{G}\right)\mathbf{H}^H $ and $ \mathbf{p}^* $ is an optimal power allocation. Note that $ \mathbf{H} = \left[\mathbf{h}_1,\dots,\mathbf{h}_k\right] $ is the collection of all downlink BS-SU channel vectors and $ \mathbf{G} = \left[\mathbf{g}_1,\dots,\mathbf{g}_M\right] $ refers to downlink BS-PU power linkage link.
\item When $(M+K-1)> n_\textsc{bs}$, $ {\rm rank} \left(\mathbf{F}_{-k}\right)=n_\textsc{bs}$ and $ \mathbf{F}_{-k}\mathbf{t}_k=\mathbf{0} $ only has the trivial solution $ \mathbf{t}_k^* =\mathbf{0}$, which implies that all SUs are deactivated. In order to avoid this effect, we will ensure that a subset $ \mathcal{S}\subset\mathcal{K} $ of active SUs is used to keep $(M+K-1)< n_\textsc{bs}$.
\end{enumerate}

Once the maximum number of users is respected, the problem is further narrowed down to an optimal power allocation problem, based on the ZF solution given by the pseudo-inverse matrix of the channel matrix:
\begin{subequations}
	\begin{align}
	\max_{\mathbf{p}\succeq 0}	\quad & \sum_{k\in \mathcal{S}} \log_2(1+\gamma_k) \\
				{\rm s.t}:	&\sum_{k\in\mathcal{S}}p_k\left|\mathbf{t}_k^*\right|^2\leq P_\textsc{bs}
	\end{align}
\end{subequations}
which solution is already known as WF solution:
\begin{equation}
p_k = \dfrac{1}{b_k}\left[{\mu} -b_k\right]^+, \quad \text{with } {\mu \,\, \text{such that \,}}  \,\sum_{k\in\mathcal{S}} \left[{\mu}-b_k\right]^+=P_\textsc{bs}
\end{equation}
where {$\mu$ is the water level,} $ b_k $ denotes the $ k$-th diagonal element of $ \left(\mathbf{H}\mathbf{H}^H\right)^{-1}${, and the operator $[\cdot]^+ = \max\{0, \cdot\}$.}

\vspace{4mm}

\section{Numerical results}
\label{Sec_Results}
Section \ref{Sec_ZFWF} presented an alternative suboptimal solution to the sum capacity initial problem, where a ZFWF manipulation eliminates the interference constraint inherent to the original optimization problem, while reducing the overal optimization problem to a power allocation strategy, which can be straightforwardly and optimally solved via water filling algorithm. 

This section analyses numerical results comparing BER and sum capacity figures-of-merit taking into account different CRN configurations. In the numerical simulations we have considered 4-QAM transmission, varying the number of PUs, SUs and $n_\textsc{bs} $, short-term fading,  while the interference from PUs to SUs was fixed to $ I_p=0$ dB. Monte Carlo simulations with $10^6$ realizations were proceeded in order to guarantee a confidence interval of 98\% and relative error of 5\% \cite{Balaban.2012}.

\subsection{Capacity comparison}
The proposed precoding technique is intended to optimize the sum-capacity of the SUs class in a MISO-CRN subject to power and interference constraints. Numerical results in this subsection are devoted to demonstrate the efficiency and effectiveness of the proposed near-optimal ZFWF precoding-based transmission design. Fig. \ref{fig_Capacity} compares the sum capacity of a CRN with different number of primary ($M$) and secondary ($K$) users and BS antennas for two power allocation strategies: the ZFWF proposed in section \ref{Sec_ZFWF} and a {ZF with} {\it equal power allocation} (ZFEP), where every antenna transmits with the same power, $P_{\rm BS}/n_{\rm BS}$. All network specifications are depicted in Table \ref{tab_results1}. In both cases the ZF strategy was used to eliminate interference from SU to others cognitive users and to PUs. Our numerical results corroborate that water filing power allocation strategy plays an important role in capacity enhancement of SUs. This fact is observed by the wide gap between ZFEP and ZFWF curves in all simulated scenarios. As expected, increasing the number of SUs or antennas at BS also reflects in sum-capacity grow, which is expected once the algorithm has optimized the sum capacity of a secondary network subject to the interference constraint. 

\begin{table}[!htbp]
	\centering
	\caption{Reference values used for simulation scenario 1}
	\begin{tabular}{ll}
		\hline
		\bf Parameter 		& \bf Value \\
		\hline \hline
		SINR 				& ${\gamma}\in[-15,\, 35] $dB\\   
		SUs 	& $ K \in\{3,\, 5,\, 10,\, 15 \} $ \\
		PUs 		& $ M \in \{1, \, 2\} $\\
		CR-BS antennas		&	$ n_\textsc{BS} \in\{8,\, 16\} $\\
		PU interference 	& $I_p = 0 $dB\\
		Modulation			& 4-QAM	\\
		\hline
	\end{tabular}
	\label{tab_results1}
\end{table}

Furthermore, increasing the number of PUs from $M=1$ to $M=2$ (Fig. \ref{fig_Capacity}.b) will decrease the sum capacity of the secondary network, once SUs have to limit their transmission to avoid degrading all primary transmissions. It is observed in red and magenta curves, which didn't present the same slope as the other curves for high SINR values in Fig. \ref{fig_Capacity}.a). \colk{This fact is also due to the interference $ I_p $ overall secondary transmission. Once there are more than one PU transmitting over the band, this fixed interference is also greater, and SUs do not have how to avoid it.}
\begin{figure}[!htbp]
	\centering
	\includegraphics[width=.8\textwidth]{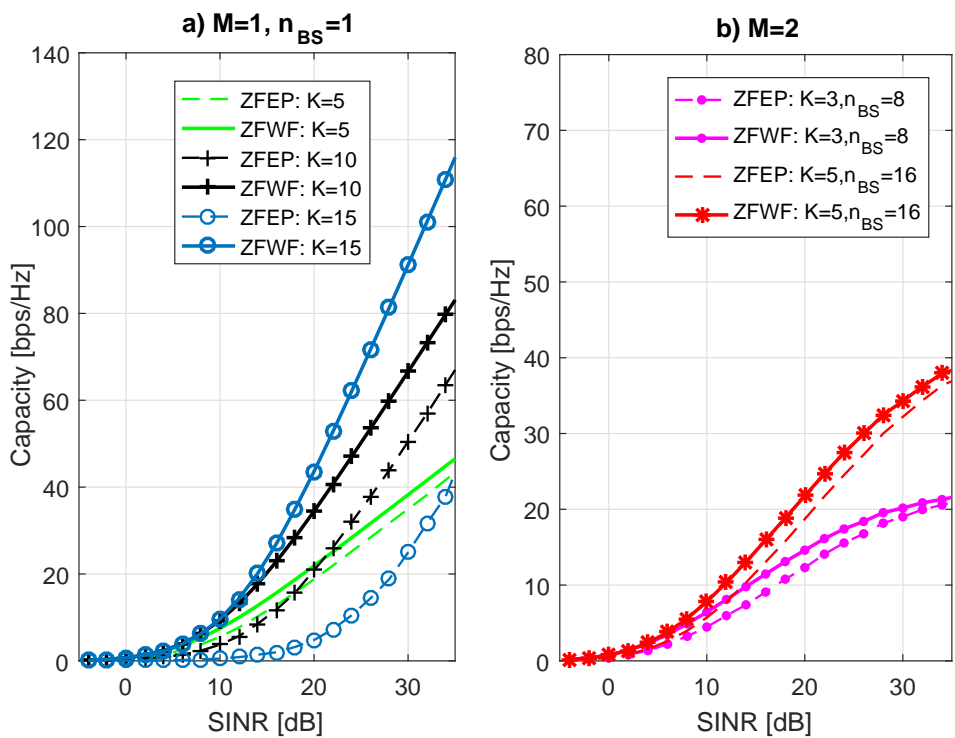}
	\vspace{-5mm}
	\caption{Sum capacity for ZFWF and ZFEP power allocation schemes.}
	\label{fig_Capacity}
\end{figure}

\subsection{Optimum Number of Secondary Users in a {CRN}}
A CR network is unique in terms of spectrum management and interference limits. As seen before, an underlay CR has strict interference thresholds and scarce transmission power. As a consequence, these constraints pose an important role in capacity of CR scenarios. {In order to achieve greater capacity, the SU has to increase} power or spacial diversity (more users or antennas at the BS). However, the presence of PUs limits this capacity enhancement. Differently from conventional MIMO systems, a CRN does not presents an unlimited increase in capacity when more users/antennas are transmitting.

In order to illustrate this phenomenon, a simulation varying $ n_\textsc{bs} $ and number of single antenna SUs was carried out for different values of SINR. The result is plotted if Fig. \ref{fig_surface} and clearly expresses the existence of an optimal point for the number of users according to each {$n_\textsc{bs}$ configuration}. This effect is explained by two major factors. Firstly, as $ K $ and $ n_{bs} $ increase, the dimension of $\mathbf{F}_{-k}$ also increases; as a consequence, there exists fewer solutions that guarantee a precoding matrix lying in the null space of $\mathbf{F}_{-k}$ and the ZF algorithm is not able to completely null the interference. Secondly, as SUs are generally low-cost, low-power radios, when $ K $ increases, an unavailable amount of transmission power is required to guarantee quality communication for all users, which ends up reducing the secondary sum capacity.
	
\begin{figure}[!htbp]
	\centering
	\includegraphics[width=.7\textwidth]{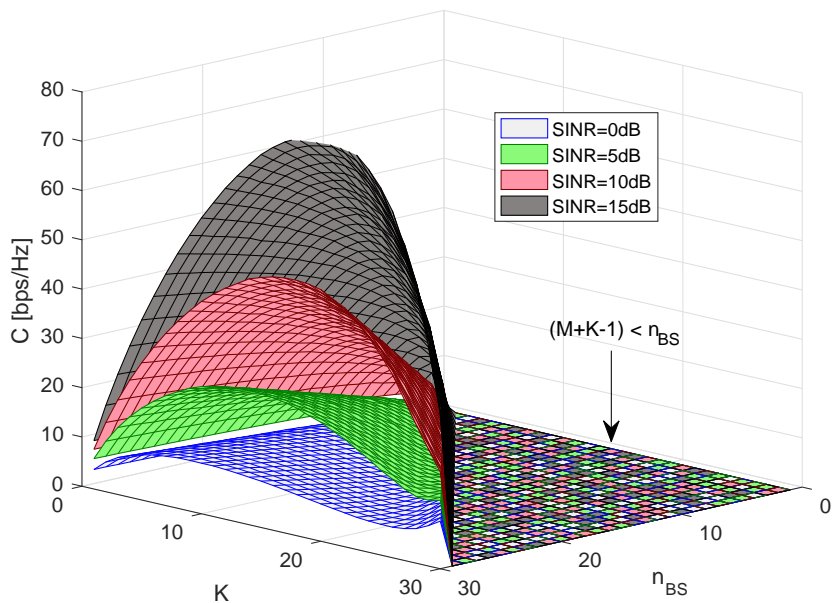}
	\vspace{-8mm}
	\caption{Capacity dependency on $ K $ and $ n_\textsc{bs} $.}
	\label{fig_surface}
\end{figure}

All cases in Fig. \ref{fig_surface} presented a peak capacity dependent on $ K $ and $ n_\textsc{bs} $. In real scenarios, the cognitive BS has a fixed number of transmitting antennas; however, it is possible to choose an adequate number of SUs aiming to maximize the sum capacity of the secondary network while guaranteeing the primary interference constraint. In order to do so, we have created a fitted model to approximate and ease the decision of how many SUs should be allowed to transmit in a certain CR network. The contour curves and maximum capacity points depicted in Fig. \ref{fig_contour} confirms a linear dependence between $ K $ and $ n_\textsc{bs} $ to achieve the maximal sum-capacity of SUs network, $ \mathcal{C}_{\max} $. Hence, the fitted curve {for the best number of users} is obtained for a specific SINR:
\begin{equation}\label{eq:fitting_@SINR}
K^* = 0.6712 \cdot n_\textsc{bs} + 0.2299, \qquad @\gamma=15 [{\rm dB}]
\end{equation}
where $ K^* $ is the number of SUs that maximizes sum capacity of a certain number of base station antennas, for an specific operating SINR network value. 
\begin{figure}[!htbp]
	\centering
	\includegraphics[width=.7\textwidth]{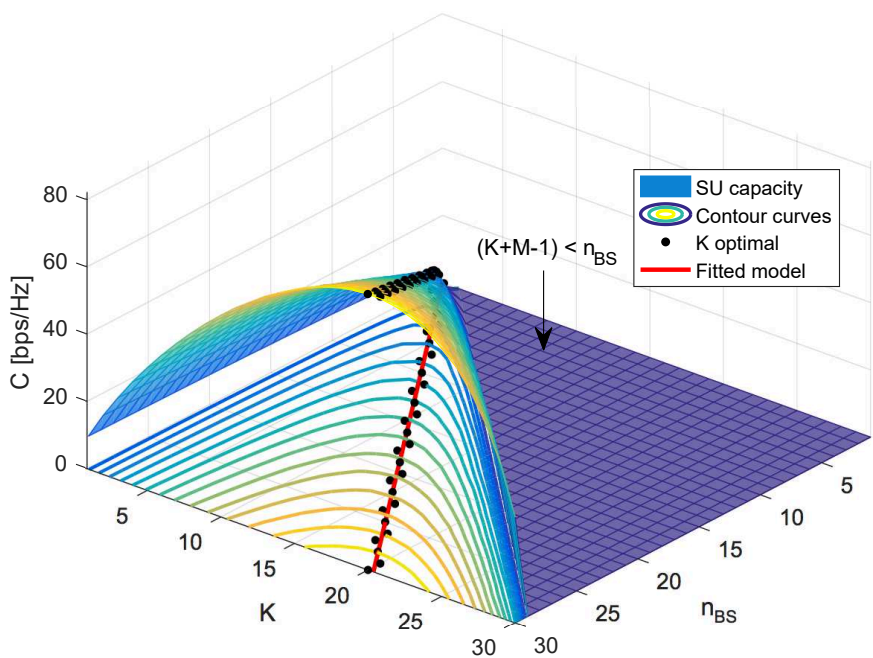}
	\vspace{-3mm}
	\caption{Sum capacity for a CR network with varying $ K $ and $ n_\textsc{bs} $ and $ \gamma=15 $dB.}
	\label{fig_contour}
\end{figure}

{Notice that all} linear fittings of $ K^* $ are dependent on SINR; as a consequence, there will exist one different equation for every desired SINR, as follows:
\begin{align*}
K^* &= 0.3071 \cdot n_\textsc{bs} + 0.5429, \qquad @\gamma=0 [{\rm dB}]\\
K^* &= 0.5357 \cdot n_\textsc{bs} + 0.3143, \qquad @\gamma=8 [{\rm dB}]\\
K^* &= 0.6893 \cdot n_\textsc{bs} + 0.2190, \qquad @\gamma=16 [{\rm dB}]\\
K^* &= 0.8143 \cdot n_\textsc{bs} - 0.0476, \qquad @\gamma=24 [{\rm dB}]
\end{align*}
Our main goal is to identify a general equation relating $ K^* $, $ n_\textsc{bs} $ and SINR. Once $ K^* $ is a linear equation {regarding the number of antennas $n_\textsc{bs}$}, we can write:
\begin{equation}
K^*(\gamma,n_\textsc{bs}) = \varphi(\gamma) \cdot n_\textsc{bs} + \beta(\gamma)
\end{equation}
where the angular coefficient ($ \varphi $) and the constant term $ \beta $ have to be estimated as a function of SINR and number of antennas. Fig. \ref{fig_fitting} illustrates the estimation method for {$\varphi$ and $\beta$}. 
\begin{figure}[!htbp]
	\centering
	\includegraphics[width=.75\textwidth]{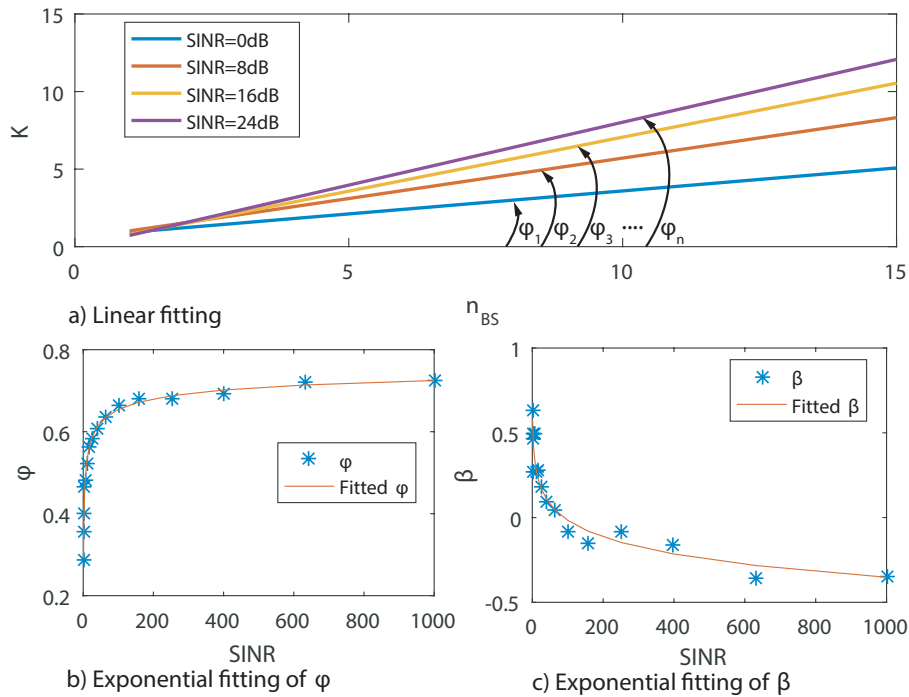}
	\vspace{-6mm}
	\caption{{Exponential Fitting procedure to estimate parameters $\varphi$ and $\beta$}.}
	\label{fig_fitting}
\end{figure}

As observed in Fig. \ref{fig_fitting} both coefficients behave log-exponentially according to SINR, which allows us to make use of an exponential fitting with SINR $\gamma$ being the independent variable:
\begin{align}
\varphi(\gamma) &= a_1 \cdot \gamma^{b_1} + c_1 \label{eq_fit_phi}\\
\beta(\gamma) &= a_2 \cdot \gamma^{b_2} + c_2 \label{eq_fit_beta}
\end{align}
By applying an exponential fitting procedure {on the data of Fig. \ref{fig_fitting}}, we are able to estimate the parameters $ a_n, \, b_n,$ and $ c_n $ of \eqref{eq_fit_phi} and \eqref{eq_fit_beta}:
\begin{center}
\begin{tabular}{ccccc}
\hline
\multicolumn{2}{c}{$\varphi(\gamma)$}  && \multicolumn{2}{c}{$\beta(\gamma)$}\\
\hline
$a_1$ &= $-0.5189$ && $a_2$ &= $-3.2938$\\
$b_1 $&= $-0.2608$ && $b_2$ &= \,\,\,\,\,$0.0360$\\
$c_1$ &= \,\,\,\,\,$0.8107$ && $c_2$ &= \,\,\,\,\,$3.8715$\\
\hline
\end{tabular}
\end{center}
\vspace{4mm}

Finally, the number of SUs that maximize the {\colk{SR} capacity of CRN for a given $ n_\textsc{bs} $ and $ \gamma$ configuration} can be {suitably} approximated by the following expression:
\begin{align}
K^*(\gamma, n_\textsc{bs}) &= \tan \left[\varphi(\gamma)\right]\cdot n_\textsc{bs} + \beta(\gamma) \nonumber\\
&= \tan\left( -0.5189 \cdot \gamma^{-0.2608} + 0.8107 \right) n_\textsc{bs}  - 3.2938  \cdot\gamma^{0.0360} + 3.8715 \label{eq_fit_K}
\end{align}

{To evaluate the consistence} of \eqref{eq_fit_K}, {Fig. \ref{fig_surface4} depicts surfaces of the sum-capacity} obtained with ZFWF algorithm and its optimum number of SUs obtained via $ \eqref{eq_fit_K} $.

\begin{figure}[h]
	\subfloat{\includegraphics[width=.5\textwidth]{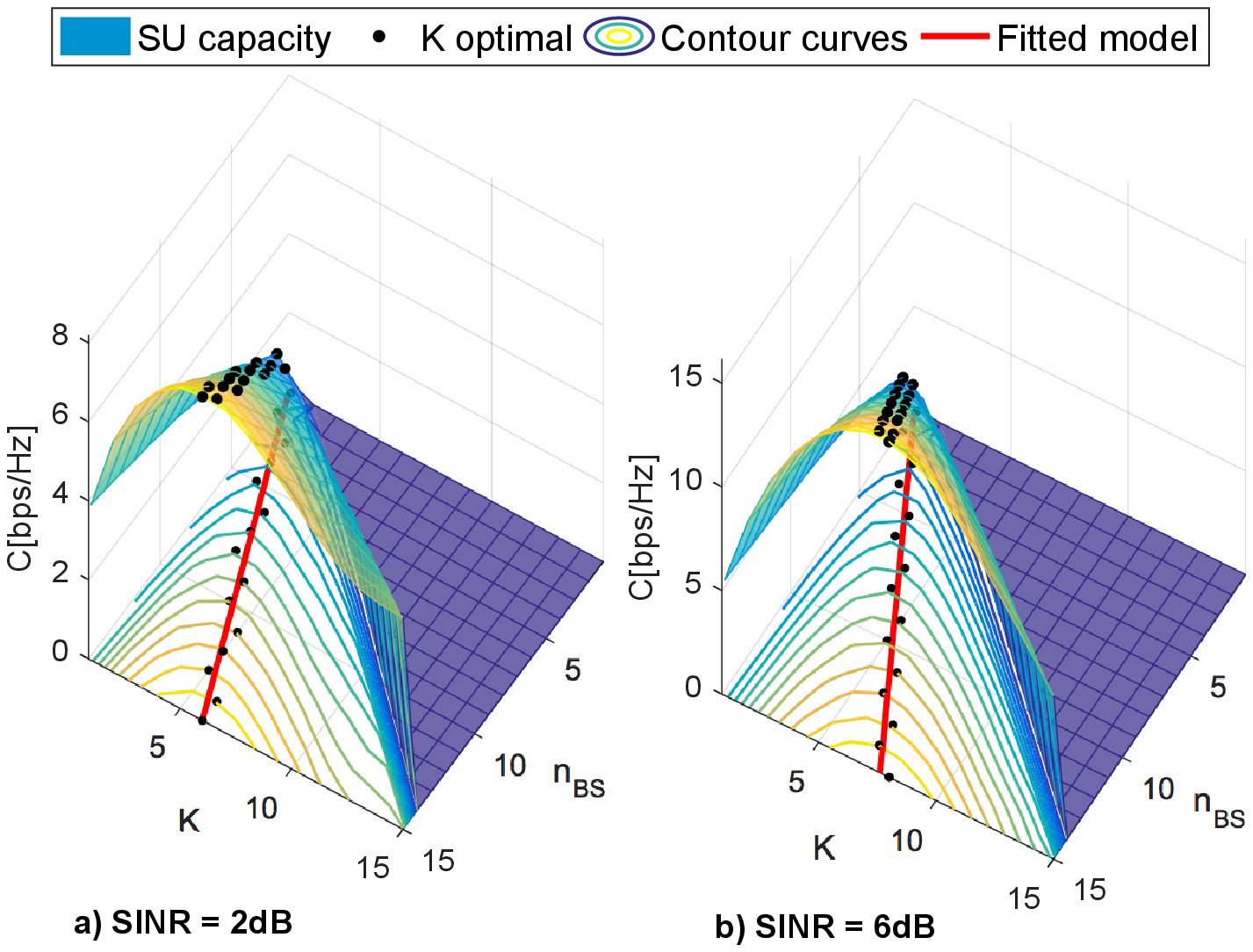}}
	\subfloat{\includegraphics[width=.5\textwidth]{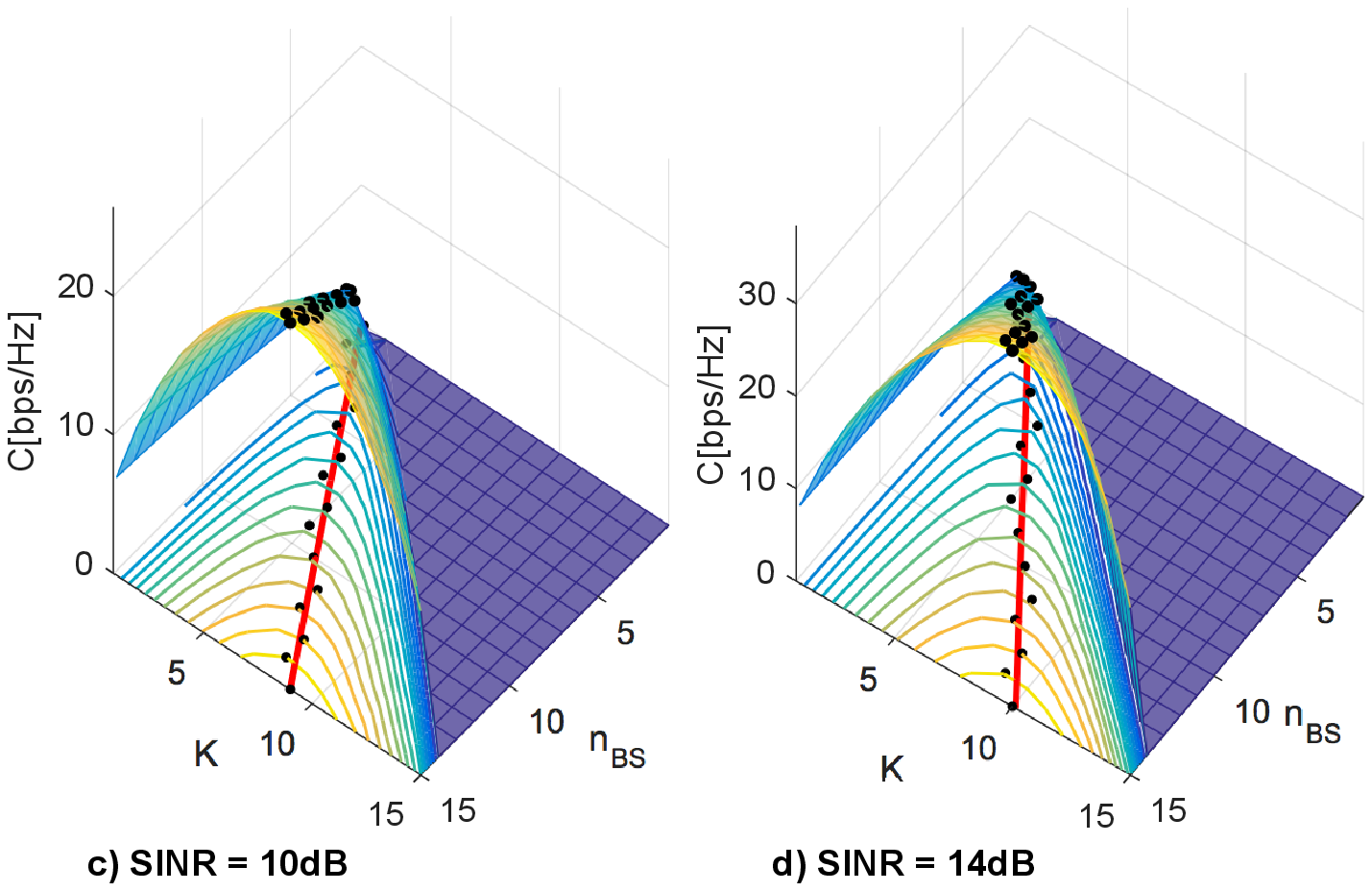}}\\[-2ex]
	\caption{{Evaluating the consistence of the proposed fitting expression for $K$ optimum}}
	\label{fig_surface4}
\end{figure}

The last analysis {aiming} to completely understand {\it how to choose the adequate number of SUs, i.e.,} {optimum number of SUs in terms of maximum achievable \colk{SR} capacity for a given SINR and number of available antennas,} {can be checked from the surface plotting of $ K^* \times  n_\textsc{bs} \times$ SINR in Fig. \ref{fig_surface2}}. The result of this subsection consists in simulating the sum capacity optimization problem as previously explained and, for every chosen $ n_\textsc{BS} $ and SINR, finding the correspondent $ K^* $ that maximizes the \colk{SR capacity in CRNs}.
\begin{figure}[!htbp]
	\centering
	\includegraphics[width=.7\textwidth]{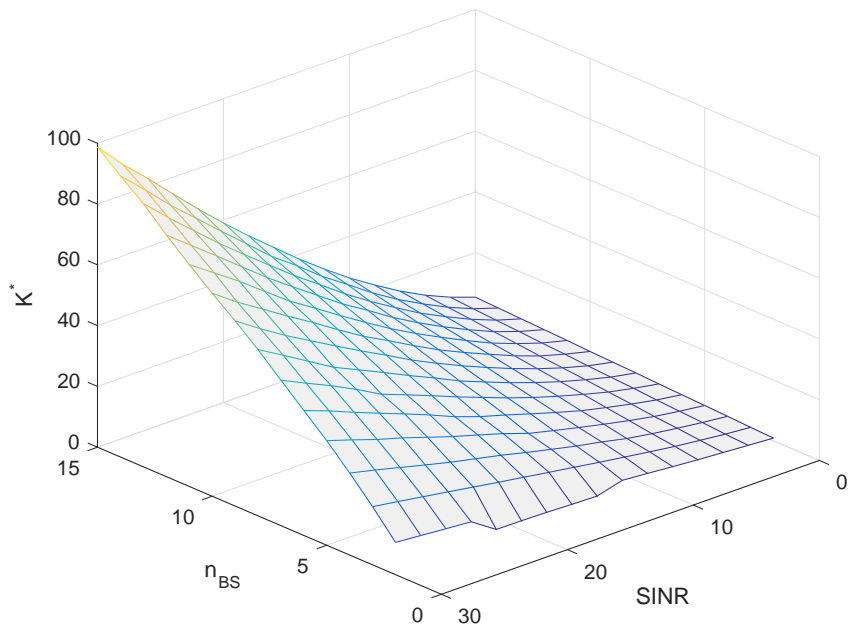}
	\vspace{-6mm}
	\caption{{Maximum \colk{SR} capacity of a CRN is achieved with the optimum number of SUs, $K^*$}.}
	\label{fig_surface2}
\end{figure}

\subsection{MMSE Precoder Comparison}
Even though ZFWF technique has proved to be efficient for capacity maximization, it is known that ZF precoding strategy usually results in  high BER figures. A strategy known due to its excellent performance in terms of BER is the MMSE precoding. Other precoding techniques have been studied to enhance performance of CRN, like MMSE-based precoders \cite{KyoungLee2011,JZhou.2008}, or even more recent precoding strategies, such as {\it bivariate probabilistic constrained programming} (BPCP) \cite{Law2017} and {\it leakage rate limiting} (LRL) precoding strategy\cite{Mohannad2016}. 

MMSE-based precoding techniques are known to present lower BER figures if compared to many other strategies. The following results fairly compare the proposed ZF technique and a MMSE-based precoder. This simulation aims to design a precoder that minimizes the  MSE for a given CRN configuration \cite{JZhou.2008} and, as a consequence, presents lower bit error rates. However, techniques in \cite{JZhou.2008,KyoungLee2011} are not created to maximize capacity, and there is no optimal power allocation in this sense. Fig. \ref{fig_Capacity_ZF_MMSE} firstly confirms that ZFWF is more efficient in terms of sum capacity than MMSE-based strategy for CR network configurations presented in Table \ref{tab_results2}.

\begin{table}[!htbp]
	\centering
	\caption{Reference values used for simulation scenario 2}
	\begin{tabular}{ll}
		\hline
		\bf Parameter 		& \bf Value \\
		\hline \hline
		SINR 				& $\gamma \in[-5,\, 35] $dB\\ 
		SUs 	& $ K \in\{10,\, 15 \} $ \\
		PUs 		& $ M = 1 $\\
		CR-BS antennas		& $ n_\textsc{BS} = 16 $\\
		PU interference 	& $I_p = 0 $dB\\
		Modulation			& 4-QAM	\\
		\hline
	\end{tabular}
	\label{tab_results2}
\end{table}

\begin{figure}[!htbp]
	\centering
	\includegraphics[width=.75\textwidth]{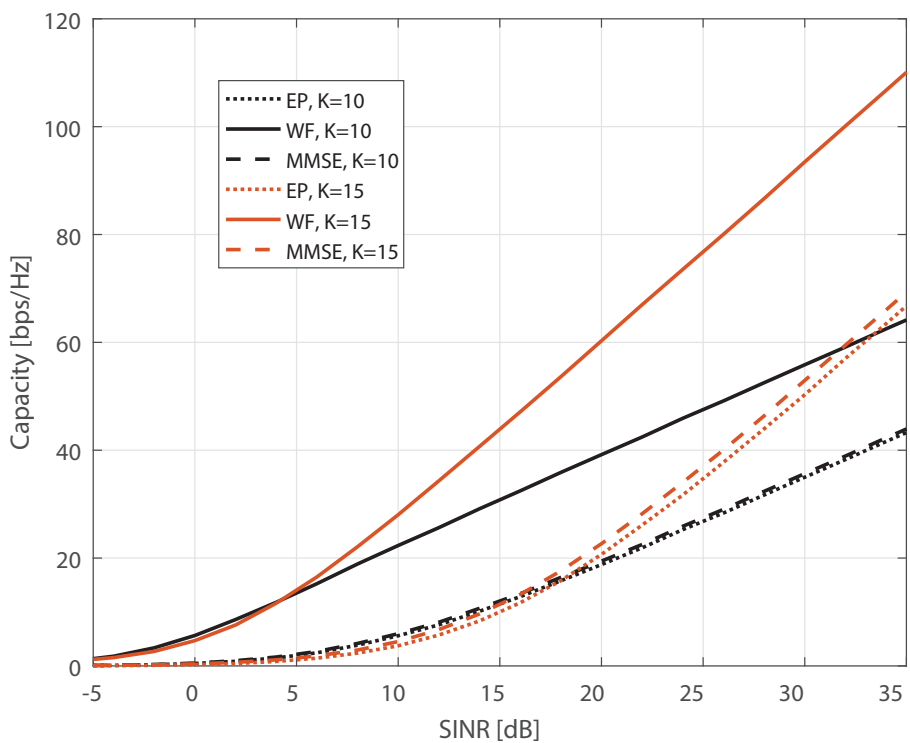}
	\vspace{-1mm}
	\caption{Sum rate capacity for ZF and MMSE beamforming techniques. $K=10$ and $K=15$ SUs.}
	\label{fig_Capacity_ZF_MMSE}
\end{figure}

As expected, the water filling algorithm -- which is a valid application for this scenario only if combined to ZF channel cancellation -- presents significant improvements in terms of sum capacity maximization, as seen in Fig. \ref{fig_Capacity_ZF_MMSE}. Specially if greater spatial diversity is exploited via increasing the number of antennas at SU-BS. The case of more SUs is also a form of increasing sum capacity. A secondary network with $K=10$ and $K=15$ SUs has been evaluated. However, increase in number of SUs also affects the BS's power limit. Additionally, more SUs in the same network end up reducing the null space in which the ZF precoding matrix must lie on, which may difficult the solution and, as SINR increases, inter user interference is also prone to increase as well. 

The greater difference from ZF to MMSE-based techniques is the capacity enhancement when water filing power allocation is applied, which is valid only if combined to ZF interference cancellation; indeed, there is not much improvement to be done in MMSE-based precoders in terms of {\colk{SR}} gain. This difference is seen in both scenario configurations of Fig. \ref{fig_Capacity_ZF_MMSE}, once both ZFEP and MMSE-based curves present almost the same results, while ZFWF shows much greater capacity.

\subsection{Bit Error Rate Comparison}
The proposed ZFWF beamforming technique is a quasi-optimal solution; also, ZF is known not to completely cancel interference is some cases. As a consequence, given the increase in capacity, some detection error may appear and BER figure-of-merit is an interesting choice to analyze the performance of a transmission system. Fig. \ref{fig_BER} presents results of BER for several system configurations. Note that, even though ZFWF was designed to optimize capacity, it also minimally affects BER, once ZFWF has slightly smaller BER for all cases. Also, increase in number of SUs or $ n_\textsc{bs} $ affects BER performance. As expected, a higher SU spatial diversity reflects in greater BER values, as seen from the overall separation between the case with $ K=15 $ users and $ n_\textsc{bs}=16$ from all other curves.

A BER floor is seen in curves with more than one PU. This is due to the fact that PUs consist of a strong interference to secondary transmissions, and this in an unavoidable matter. Once PUs have priority in any CR transmission, if one or more PUs wish to start transmitting over a certain frequency, SUs just have to learn how to deal with it. In order to do so, SUs must limit its transmission power and try to filter out PUs's signals. As a consequence, the extra power needed to keep low BER levels has to be controlled and BER floors unavoidably appear. However, active interference cancellation techniques can be used if lower BER values are required for networks with more than one PU.
\begin{figure}[!htbp]
	\centering
	\includegraphics[width=.75\textwidth]{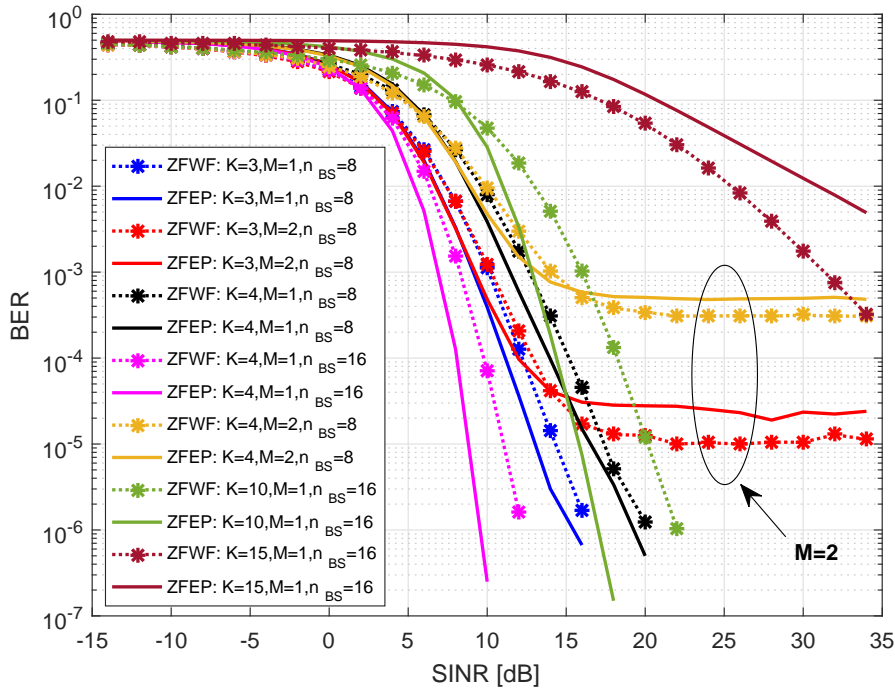}
	\vspace{-1mm}
	\caption{BER for ZFWF and ZFEP power allocation schemes {with $M=1$ and $M=2$ PUs.}}
	\label{fig_BER}
\end{figure}

\section{Final Remarks}
\label{Sec_Conclusion}
This article was firstly intended to present a consistent mathematical demonstration of convexity analysis for the \colk{SR} maximization problem of broadcast (DL) underlay MISO cognitive networks. We have applied the ZFWF as a sub-optimal solution to maximize the SUs' capacity while minimally interfering on primary transmissions. Our numerical results firstly compared and widely corroborated the superiority of the proposed beamforming technique regarding the ZF combined to the equal power allocation approach. Even though the precoding was designed to maximize sum capacity, numerical results demonstrated that both power allocation strategies, when applied to some CR-MISO scenario, result in similar performance; however, results are more expressive when greater spatial diversity is employed. Under higher spatial diversity scenarios, the interference plays such an important role in CRNs capacity and BER performance; indeed, even with optimal power allocation techniques, the capacity is very limited by the interference term in SINR expression. Numerical results for bit error rate have evidenced that, for this case, sum capacity optimization implies in BER performance loss.

Comparison ZFWF, ZFWP and MMSE precoding techniques showed that, power allocation alone brings some benefits to overall network capacity, but BER is still strongly affected, depending on the system SU and PU configuration. BER performance results for all simulated cases indicated that, apart from highlighting which technique presents better results, an expressive BER floor is seen when more than one PU is present (increasing and unavoidable interference). This fact is explained not only by the power limit constraint imposed to the sum capacity optimization problem but also by the increase in interference caused by PUs, once this is an unavoidable matter.

The comparison of ZF and MMSE-based precoding techniques {has confirmed that a} great advantage is obtained when ZF interference cancellation is applied: the possibility of dealing with independent channels and, consequently, application of water filing power allocation to achieve much greater {\colk{SR}} for a given secondary network configuration.

An important result {unveiled} in this article is the linear dependence between $ K $ and $ n_\textsc{bs} $ to achieve maximum {\colk{SR} for SUs in a} MU-MISO network. The approximation here suggested has a rooted MSE equals to 0.29, which gives us a fair estimation of the optimal number of SUs for a given architecture of CR-BS. Also, an extended dependency between $ K^* $ and SINR was {established. Our numerical} results allowed us to {propose} an exponential approximation of $ \varphi $ and $ \beta $ to achieve an expression relating $ K^* $, $ n_\textsc{bs} $ and SINR. This result offers a simple {and effective procedure} to find the optimal number of SUs that can be allocated to a certain cognitive {radio} network.

\section*{Acknowledgement}
This work was supported in part by the National Council for Scientific and Technological Development (CNPq) of Brazil under Grant 304066/2015-0.

\end{document}